\documentclass[conference]{IEEEtran}
\IEEEoverridecommandlockouts
\usepackage{cite}
\usepackage{amsmath,amssymb,amsfonts}
\usepackage{graphicx}
\usepackage{textcomp}
\usepackage{xcolor}
\def\BibTeX{{\rm B\kern-.05em{\sc i\kern-.025em b}\kern-.08em
    T\kern-.1667em\lower.7ex\hbox{E}\kern-.125emX}}

\usepackage{subfigure}
\usepackage[noend]{algpseudocode}
\usepackage{enumitem}
\usepackage{booktabs} 
\usepackage{balance}
\usepackage{algorithm}
\usepackage{subfigure}
\usepackage[noend]{algpseudocode}
\usepackage{enumitem}
\usepackage{hyperref}
\usepackage{comment}
\usepackage{bm}
\usepackage{amsmath}
\DeclareMathOperator\arctanh{arctanh}
\DeclareMathOperator\arcosh{arcosh}

\begin{document}


\title{Time-aware Hyperbolic Graph Attention Network for
Session-based Recommendation}



\author{
    \IEEEauthorblockN{Xiaohan Li\IEEEauthorrefmark{1}\IEEEauthorrefmark{2}, Yuqing Liu\IEEEauthorrefmark{1}\IEEEauthorrefmark{3}, Zheng Liu\IEEEauthorrefmark{3}, Philip S. Yu\IEEEauthorrefmark{3}}
    \IEEEauthorblockA{\IEEEauthorrefmark{2}Walmart Global Tech, Sunnyvale, CA, USA
    \\xiaohan.li@walmart.com}
    \IEEEauthorblockA{\IEEEauthorrefmark{3}University of Illinois at Chicago, Chicago, IL, USA
    \\\{yliu363, zliu212, psyu\}@uic.edu}
\thanks{\IEEEauthorrefmark{1}Both authors contributed equally to this research.}    
}

\maketitle

\begin{abstract}
Session-based Recommendation (SBR) is to predict users' next interested items based on their previous browsing sessions. Existing methods model sessions as graphs or sequences to estimate user interests based on their interacted items to make recommendations. In recent years, graph-based methods have achieved outstanding performance on SBR. However, none of these methods consider temporal information, which is a crucial feature in SBR as it indicates timeliness or currency. Besides, the session graphs exhibit a hierarchical structure and are demonstrated to be suitable in hyperbolic geometry. But few papers design the models in hyperbolic spaces and this direction is still under exploration.

In this paper, we propose \textbf{Time-aware Hyperbolic Graph Attention Network} (TA-HGAT) --- a novel hyperbolic graph neural network framework to build a session-based recommendation model considering temporal information. More specifically, there are three components in TA-HGAT. First, a hyperbolic projection module transforms the item features into hyperbolic space. Second, the time-aware graph attention module models time intervals between items and the users' current interests. Third, an evolutionary loss at the end of the model provides an accurate prediction of the recommended item based on the given timestamp. TA-HGAT is built in a hyperbolic space to learn the hierarchical structure of session graphs. Experimental results show that the proposed TA-HGAT has the best performance compared to ten baseline models on two real-world datasets.
 
\end{abstract}





\section{Introduction}
Recommender systems have been an effective solution to help users overcome the information overload on the Internet. Many applications are developed based on this rationale, including online retail~\cite{zhou2018deep}, music streaming~\cite{van2013deep}, and content sharing~\cite{kumar2019predicting}. To better understand users, modeling their browsing sessions is a useful solution as sessions indicate their current interests. Session-based recommendation (SBR) predicts the users' next interested items by modeling users' sessions. Deep learning models, including Recurrent Neural Networks (RNNs)~\cite{hidasi2015session, li2017neural}, Memory Networks~\cite{liu2018stamp}, and Graph Neural Networks (GNNs)~\cite{wu2019session, chen2020handling} are applied to this problem and have achieved state-of-the-art performance.

Recently, the most influential works on dealing with SBR are GNN-based methods. The GNN-based methods~\cite{wu2019session, chen2020handling, wang2020global, xu2019graph, qiu2019rethinking, yu2020tagnn} take each session as a graph to learn the items' internal relationship and their complex transitions. The most representative model is SR-GNN \cite{wu2019session}, which is the first work to apply GNN on session-based recommendation and achieve state-of-the-art performance. Based on SR-GNN, \cite{xu2019graph, yu2020tagnn} improve SR-GNN with attention layers. \cite{qiu2019rethinking, chen2020handling} consider the item order in the session graph to build the models. \cite{wang2020global, liu2021case4sr, pang2022heterogeneous} take additional information such as global item relationship, item categories, and user representations into account to devise more extensive models. HCGR \cite{guo2021hcgr} models session graphs into a hyperbolic space to extract hierarchical information.

Although the existing GNN-based methods have achieved satisfactory performance, they still suffer from two limitations.
\textit{First}, unlike sequence-based models, graph structure cannot explicitly show the temporal information between items. Time interval is a crucial feature and can significantly improve the recommendation performance \cite{li2020time, ye2020time}, but it is ignored in the existing graph-based SBR models. Moreover, though modeling sessions into graphs has the advantage of learning items complex transitions \cite{wu2019session}, the sequential relation between items is unclear in the session graph because the beginning and end of a session are ambiguous under the graph structure. \textit{Second}, according to \cite{wilson2014spherical, bronstein2017geometric}, graph data exhibits an underlying non-Euclidean structure, and therefore, learning such geometry in Euclidean spaces is not a proper choice. As a result, some recent studies \cite{wang2021fully, guo2021hcgr, sun2021hgcf} reveal that the real-world datasets of recommender systems usually exhibit tree-like hierarchical structures, and hyperbolic spaces can effectively capture such hierarchical information. Therefore, it is worth trying to learn session graphs in hyperbolic spaces.

Hyperbolic spaces have the ability to model hierarchical structure data because they expand faster than Euclidean spaces. They can expand exponentially, but Euclidean spaces only expand polynomially. Existing work \cite{guo2021hcgr} demonstrates the hierarchical structure of session graphs. However, modeling session graphs in hyperbolic spaces is still under exploration. First, 
time intervals indicate the correlation between two items. Since hyperbolic embedding is a better match to session graphs, it is necessary to define a new framework to identify the time intervals in the edges of session graphs. Second, learning users' current interests in the graph is crucial, but it is difficult to realize in hyperbolic spaces. Previous works \cite{liu2018stamp, wu2019session} devise models in Euclidean space based on the last item in the session. The last item plays an important role in predicting the next item because it represents the users' current interest. However, it is more challenging to model this feature in a hyperbolic space as the operations in hyperbolic spaces are more complicated than the Euclidean space. Third, when taking the time information into consideration, we can not only make next-item recommendations, but also provide recommendations based on a specific timestamp.

To tackle the above challenges, we propose Time-aware Hyperbolic Graph Attention Network (TA-HGAT), a hyperbolic GNN considering the comprehensive time-relevant features. Specifically, we project the item's original features into a Poincar\'e ball space via a hyperbolic projection layer. Then, we design a time-aware hyperbolic attention mechanism to learn the time intervals and users' current interests together in a hyperbolic space. It includes two modules: hyperbolic self-attention with time intervals and hyperbolic soft-attention with users' current interests. Finally, the model is trained via an evolutionary loss to predict which item the user may be interested in at a specific timestamp. All these three components are based on a fully hyperbolic graph neural network framework.

Here, we summarize our contributions as follows:
\begin{itemize}
    \item To the best of our knowledge, this is the first paper that models temporal information in a hyperbolic space to improve the performance of the recommender system. We go beyond the conventional Euclidean machine learning models to model users’ time-relevant features in a more delicate manner.
    \item We propose TA-HGAT, a hyperbolic GNN-based framework with three main components: hyperbolic projection, time-aware hyperbolic attention, and evolutionary loss. These three components work together in an end-to-end GNN to model items' time intervals and users' current interests. In the end, our model provides a time-specific recommendation.
    \item We conduct experiments on two real-world datasets and compare our model with ten baseline models. The experiment results demonstrate the effectiveness of the TA-HGAT in MRR and Precision.
\end{itemize}


\section{Preliminary}
\subsection{Graph neural network}
\label{GNN}
GNNs \cite{kipf2016semi, hamilton2017inductive} are designed to handle the structural graph data. In GNNs, aggregation is the core operation to extract structural knowledge. By aggregating neighboring information, the central node can gain knowledge from its neighbors passed through edges and learn the node embedding. GNNs have been demonstrated to be powerful in learning node embeddings, so they are widely used on many node-related tasks such as node classification\cite{hamilton2017inductive}, graph classification\cite{ying2018hierarchical}, and link prediction\cite{kipf2016semi}.

Based on the aggregation operation, the forward propagation of a GNN on graph $G=(\mathcal{V},\mathcal{E})$ is to learn the embedding of node $v_i \in \mathcal{V}$ via aggregating its neighboring nodes.
We suppose that the initial node embedding of each node $i$ is $\mathbf{h}^{(0)}_i$, which generally is the feature of the node. In each hidden layer of a GNN, the embedding of the central node $\mathbf{h}^{(l)}_i$ is learned from the aggregated embedding of the neighboring nodes in the previous hidden layer $\mathbf{h}^{(l-1)}_i$. The process is described in math as follows:
\begin{equation}
    \mathbf{h}^{(l)}_i = \sigma \Big(\mathbf{W}^{(l)} ( \underset{j \in \mathcal{N}_i}{\text {AGG}}(\mathbf{h}^{(l-1)}_j)\Big),
\end{equation}

where $\mathcal{N}_i$ represents the set of all neighbors of node $i$ in the graph, including the node $i$ itself. 
The aggregation function $\text{AGG}(\cdot)$  integrates the neighboring information together. A non-linear activation function $\sigma$, e.g., sigmoid or LeakyReLU, is applied to generate the embedding of node $i$ in the layer $l$.  

Based on the vanilla GNN we mentioned above, GAT \cite{velivckovic2017graph} is proposed to improve GNNs with self-attention mechanism \cite{vaswani2017attention}. Specifically, 
for all the neighbors of node $i$, we need to learn the attention coefficients for all its neighbors to calculate the importance of each neighbor node in the aggregation. Suppose the attention coefficient of the node pair $(i,j)$ is $\alpha_{ij}$, the process of learning $\alpha_{ij}$ is
\begin{equation}
    \alpha_{ij} = softmax(\mathbf{d}_{ij}) = \frac{\exp(\mathbf{d}_{ij})}{\sum_{k \in \mathcal{N}_i} \exp(\mathbf{d}_{ik}) },
\end{equation}
where $\mathbf{d}_{ij}$ is the correlation between node $i$ and $j$. $\mathbf{d}_{ij}$ here can be the joint embeddings of node $i$ and $j$, e.g., concatenation of node embeddings or similarity of the node pair.

\subsection{Hyperbolic spaces}
In definition, hyperbolic space is a homogeneous space with negative curvature. It is a smooth Riemannian manifold, which can be modeled in several hyperbolic geometric models, including Poincar\'e ball model\cite{nickel2017poincare}, Klein model\cite{gulcehre2018hyperbolic}, Lorentz model\cite{nickel2018learning}, etc. In this paper, we choose the Poincar\'e ball model because the distance between two points grows exponentially, which fits well with the hierarchical structure of the session graph. Formally, the space of the $d$-dimensional Poincar\'e ball $\mathbb{P}_c^d$ is defined as
\begin{equation}
    \mathcal{P}_c^d = \{\mathbf{x} \in \mathbb{R}^d, c\|\mathbf{x}\| \textless 1\},
\end{equation}
where $c$ is the radius of the ball and $\mathbf{x}$ is any point in manifold $\mathcal{P}$. If $c = 0$, then $\mathbb{P}_c^d = \mathbb{R}^d$ and the ball is equal to the Euclidean surface. In this paper, we set $c = 1$. The tangent space $\mathcal{T}_{\mathbf{x}}\mathcal{P}$ is a $d$-dimensional vector space approximating $\mathcal{P}$ around $\mathbf{x}$, which is isomorphic to the Euclidean space. With the exponential map, a vector in the Euclidean space can be mapped to the hyperbolic space. The logarithmic map is the inverse of the exponential map, which projects the vector back to the Euclidean space.

In hyperbolic spaces, the fundamental mathematical operations of neural networks (e.g., addition and multiplication) are different from those in Euclidean space. In this paper, we choose M\"obius transformation as an algebraic operation for studying hyperbolic geometry. For a pair of random vectors $(\mathbf{a}, \mathbf{b})$, we list the operations that will be used in our model as follows:
\begin{itemize}
    \item M\"obius addition $\oplus$ \cite{liu2019hyperbolic} is to perform addition operation of $\mathbf{a}$ and $\mathbf{b}$.
    \begin{equation}
    \mathbf{a} \oplus \mathbf{b} = \frac{(1+2 \langle \mathbf{a}, \mathbf{b} \rangle + \| \mathbf{b} \|^2)\mathbf{a} + (1- \| \mathbf{a} \|^2)\mathbf{b}}{1 + 2 \langle \mathbf{a}, \mathbf{b} \rangle + \| \mathbf{a} \|^2\| \mathbf{b} \|^2}.
    \label{hyper_add}
\end{equation}
    \item M\"obius matrix-vector multiplication $\otimes$ \cite{ganea2018hyperbolic} is employed to transform $\mathbf{a}$ with matrix $\mathbf{W}$.
\begin{equation}
    \mathbf{W} \otimes \mathbf{a} =  \tanh(\frac{\| \mathbf{W} \space \mathbf{a} \|}{\| \mathbf{a} \|}\tanh^{-1}(\| \mathbf{a} \|)),
\label{vector_multiply}
\end{equation}
    \item M\"obius scalar multiplication $\otimes$ is the multiplication of a scalar $\alpha$ with a vector $\mathbf{b}$.
\begin{equation}
    \alpha \otimes \mathbf{b}= \tanh(\alpha \tanh^{-1}(\| \mathbf{b} \|))\frac{\mathbf{b}}{\| \mathbf{b} \|}
    \label{scalar_multiply}
\end{equation}
    \item Exponential map transforms $\mathbf{a}$ from the Euclidean space to a chosen point $\mathbf{x}$ in a hyperbolic space.
    \begin{equation}
    \exp_\mathbf{x}(\mathbf{a}) = \mathbf{x} \oplus (\tanh(\frac{\lambda_\mathbf{x} \| \mathbf{a} \| }{2}) \frac{\mathbf{a}}{\| \mathbf{a} \|}),
\label{exp_map}
\end{equation}
    \item Logarithmic map projects the vector $\mathbf{a}$ back to the Euclidean space.
    \begin{equation}
        \log_\mathbf{x}(\mathbf{a}) = \frac{2}{\lambda_\mathbf{x}}\arctanh(\|-\mathbf{x} \oplus \mathbf{a} \|) \frac{-\mathbf{x} \oplus \mathbf{a}}{\|-\mathbf{x} \oplus \mathbf{a} \|}
    \label{log_map}
    \end{equation}
    \item $\lambda_\mathbf{x}$ is the conformal factor.
\begin{equation}
    \lambda_\mathbf{x} = \frac{2}{1 - \| \mathbf{x} \|^2}.
    \label{lambda}
\end{equation}
\end{itemize}

\section{Model}
\begin{figure*}[!htbp]
\centering
\includegraphics[scale=0.4]{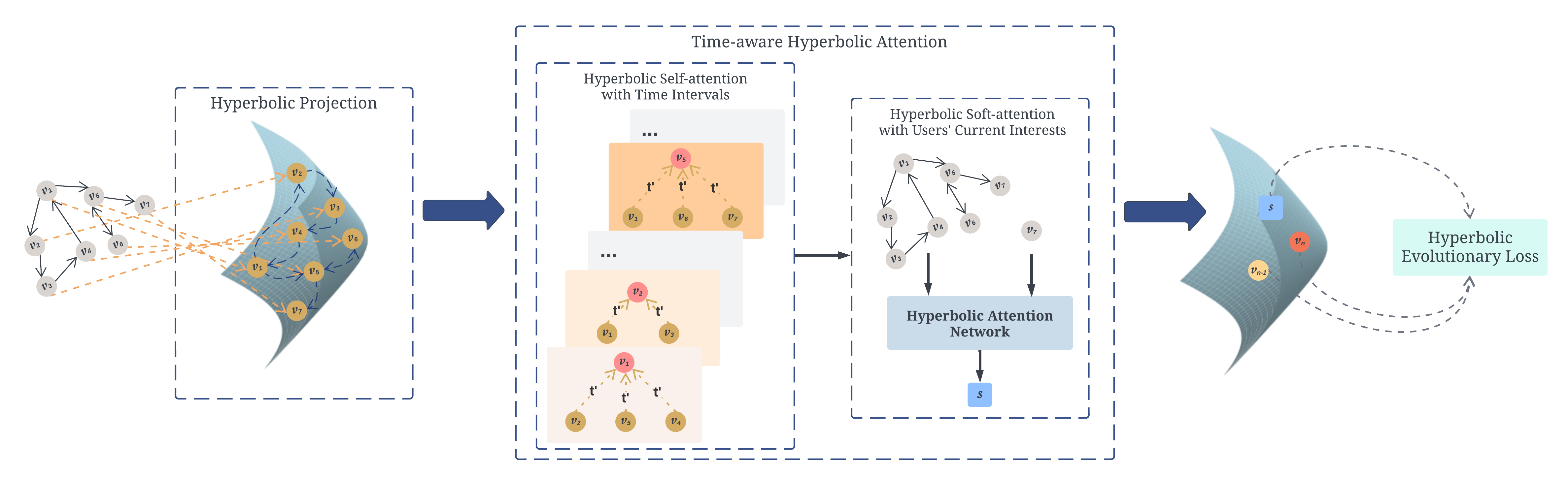}
\caption{Illustration of TA-HGAT. First, it builds directed session graphs based on the session sequences, and then projects the embeddings from the Euclidean space to the hyperbolic space. Next, hyperbolic self-attention is adopted to aggregate neighboring information and time intervals $t'$. After that, each session graph is represented as a session embedding using a hyperbolic soft-attention mechanism. Finally, TA-HGAT predicts top-$k$ items that are most likely to be clicked at the next timestamp for each session.}
\label{framework}
\end{figure*}
In this section, we present the framework of our proposed Time-aware Hyperbolic Graph Attention Network (TA-HGAT), which is designed to model the temporal information in the hyperbolic session graph. First, we define the session-based recommendation task. Then we illustrate the three main components of the model: hyperbolic projection, time-aware hyperbolic attention, and hyperbolic evolutionary loss. These three components train the model with time-relevant features and provide the recommendation results given a specific timestamp. The overall structure of TA-HGAT is shown in Figure\ref{framework}.

\subsection{Problem definition}
Session-based recommendation (SBR) is to predict the item a user will click next based on the user-item interaction sessions. Generally, it models the user’s short-term browsing session data to learn the user's current interest. Here we formulate the SBR problem mathematically as below.

In the SBR problem, a session is denoted as $S = \{v_1, v_2, \cdot\cdot\cdot, v_n\}$ ordered by timestamps. Each $v$ in $S$ is an item, and the item set is $V_s$, which consists of all unique items in this session. To model the session into a directed graph, we take all items as nodes and the item-item sequential dependency as the edges to construct the session graph. The graph is denoted as $G_s = (V_s, E_s)$, where $V_s, E_s$ are the node and edge sets, respectively. Each edge connects two consecutive items, which is formulated as $e = (v_{t-1}, v_t)$. Our target is to learn the embeddings of items and the session and generate the ranking of the items that the user may be interested in at the next timestamp.

\subsection{Hyperbolic projection}
In GNN, each node needs input as the initial embedding. Accommodated to SBR, the input of a GNN is the feature of items such as category or description. The initial embedding of item $i$ is $\mathbf{h}_i^0$. However, most feature embedding methods are based on the Euclidean space. To make the item features available in the hyperbolic space, we use the exponential map defined in Eq. \ref{exp_map} to project the initial item embeddings to the hyperbolic space. Specifically, the projection process is formulated as
\begin{equation}
    \mathbf{m}_i = \exp_\mathbf{x}(\mathbf{h}_i^0),
\label{initial}
\end{equation}
where $\mathbf{m}_i$ is the mapped embedding in the hyperbolic space and $\mathbf{x}$ is the chosen point in the tangent space.

To achieve a high-level latent representation of the node features, we also add a linear transformation parameterized by a weight matrix $\mathbf{W}_1 \in \mathbb{R}^{d' \times d}$, where $d'$ is the dimension of $\mathbf{m}_i$ and $d$ is the dimension of the node's final embedding. Please note that $\mathbf{W}_1$ is a shared weight matrix for all nodes. M\"obius matrix-vector multiplication defined in Eq. \ref{hyper_add} is employed to transform $\mathbf{m}_i$ and the process is
\begin{equation}
    \mathbf{h}_i^1 = \mathbf{W}_1 \otimes \mathbf{m}_i,
\label{vector_multiply}
\end{equation}
where $\mathbf{h}_i^1$ is the transformed embedding, which is also used as the initial node embedding in the following steps.

\subsection{Time-aware hyperbolic attention}
According to \cite{guo2021hcgr, wang2021fully, xu2020product, li2021hyperbolic}, embedding users and items in hyperbolic spaces is a significant improvement of graph-based recommender systems. However, none of these works model the time intervals and users' current interests in hyperbolic spaces. Our proposed model TA-HGAT is the first attempt to solve the problem, in which time-aware hyperbolic attention is the core component. It is composed of two attention layers: 1) Hyperbolic self-attention in the aggregation process, which considers time intervals between items; 2) Hyperbolic soft-attention in the session embedding learning, which models the user's current interest.

\subsubsection{Hyperbolic self-attention with time intervals}
According to Section \ref{GNN}, a key step in graph attention is to learn the attention coefficient $\alpha_{ij}$ for each node pair $(i,j)$. $\alpha_{ij}$ means the importance of the neighbors to the central node. To learn the $\alpha_{ij}$, unlike the traditional attention networks which apply linear transformation \cite{velivckovic2017graph} or inner product\cite{vaswani2017attention}, here we use the distance of the node embeddings in the hyperbolic space. Specifically, we denote the distance of node pair $(i,j)$ as $(\mathbf{h}_i, \mathbf{h}_j)$, which is calculated as
\begin{equation}
    d(\mathbf{h}_i^l, \mathbf{h}_j^l) = \arcosh (1+2\frac{\| \mathbf{h}_i^l - \mathbf{h}_j^l \|^2}{(1-\| \mathbf{h}_i^l \|^2)(1-\| \mathbf{h}_j^l \|^2)}).
\label{distance}
\end{equation}
Then with the node distances, we further learn the attention coefficient $\alpha_{ij}$ of node $i$ with all its neighbors (including itself) $\mathcal{N}_i$ as
\begin{equation}
    \alpha_{ij} = softmax(d_{ij}) = \frac{\exp(d_{ij})}{\sum_{k \in \mathcal{N}_i} \exp(d_{ik}) },
\end{equation}

The reason that we use distance in the hyperbolic space to calculate attention coefficients is because of two advantages. First, attention coefficients in Euclidean spaces are usually calculated by linear transformation \cite{velivckovic2017graph} or inner product\cite{vaswani2017attention}, which fail to meet the triangle inequality. In hyperbolic space, the learned attention coefficients are able to meet this criterion and preserve the transitivity among nodes. Second, the attention coefficient of the node $i$ with itself is
$\alpha_{ii} = d(\mathbf{h}_i, \mathbf{h}_i) = 0$, so the effect of the central node itself will not affect the calculation of attention coefficients.

After we achieve attention coefficients, the next step is to aggregate the node embeddings to learn the central node embedding of the next layer. Here the learned attention coefficients serve as the weights applied to the embeddings of neighbor nodes. The process is formulated as
\begin{equation}
    \mathbf{h}_i^{l+1} = \sigma (\sum_{j \in \mathcal{N}_{i}}^\oplus \alpha_{ij} \otimes \mathbf{h}_j^l), 
\label{aggregation1}
\end{equation}
where $\sum^\oplus$ is the M\"obius addition of the weighted neighbor node embeddings and $\sigma$ is a nonlinear function such as sigmoid and LeakyReLU. Different from Eq. \ref{vector_multiply}, the $\otimes$ in Eq. \ref{aggregation1} is M\"obius scalar multiplication defined in Eq. \ref{scalar_multiply}.

To integrate the temporal information into the attention layer, the core idea is to incorporate the time intervals into the aggregation process. Specifically, we transform the time intervals to the vectors in the hyperbolic space and combine the time vectors with the neighbor node embeddings for aggregation. As time intervals are continuous values, we project the time interval values into vectors with a mapping function. The mapping process is 
\begin{equation}
    \mathbf{h}_{t'} = \mathbf{w}_t \otimes (t^+-t),
\label{interval}
\end{equation}
where $t'=t^+-t$ is the time interval, $\otimes$ here is M\"obius matrix-vector multiplication, and $\mathbf{w}_t$ is the transition vector to project the time interval to a vector. In this paper, if two items have multiple time intervals between them, we choose the closest one. This process is done in the data preprocessing part before modeling.

Motivated by TransE \cite{bordes2013translating}, time-aware hyperbolic attention translates the neighbor node embedding to the central node embedding via temporal information, so the joint embedding of nodes embedding and time embedding is generated by M\"obius addition, which is represented as $\mathbf{h}_j^l \oplus \mathbf{h}_{t'}$. 

In Eq. \ref{aggregation1}, all neighbors of the central node $i$ are aggregated by M\"obius addition. As the M\"obius addition is complicated and consumes more computation resources than the addition in the Euclidean space, here we simplify the calculation in Eq. \ref{aggregation1} using the logarithmic map to project the embeddings into a tangent
space (Euclidean space) to conduct aggregation operation. Then the embeddings are projected back to the hyperbolic manifold with the exponential map.
Therefore, we can re-write the aggregation process in Eq. \ref{aggregation1} as
\begin{equation}
    \mathbf{h}_i^{l+1} = \exp \bigg( \sigma \Big(\sum_{j \in \mathcal{N}_{i}} \log( \alpha_{ij} \otimes (\mathbf{h}_j^l \oplus \mathbf{h}_{t'}) )\Big) \bigg).
\label{aggregation2}
\end{equation}

\subsubsection{Hyperbolic soft-attention with users' current interests}
In the process above, we update the embedding of node $i$ with its neighbors and time intervals. To make recommendations based on the learned node embeddings, we also need to know the global embedding of the session graph by aggregating all node embeddings. Instead of simply adding all node embeddings together, we also provide another solution to learn the graph embedding while considering users' current interests based on the most recent interacted items.

Understanding users' current interests are one of the main tasks in SBR. In the previous studies \cite{liu2018stamp, wu2019session, yu2020tagnn}, the last item in the session is the most related feature in this task. To learn from the correlation of the last item $p$ with each of the other items in the session, we adopt a soft-attention mechanism to generate attention coefficients for item $p$ with all other items, which represent the importance of items w.r.t. the current timestamp. The learning process of the global session embedding $\mathbf{h}_s$ is
\begin{equation}
    \beta_{pq} = \mathbf{x}^\intercal \otimes \sigma \Big(\mathbf{W}_2 \otimes \mathbf{h}_p) \oplus (\mathbf{W}_3 \otimes \mathbf{h}_q) \oplus \mathbf{c}\Big),
\end{equation}
\begin{equation}
    \mathbf{h}_s = \exp \bigg( \sigma \Big(\sum_{q \in V_s} \log ( \beta_{pq} \otimes \mathbf{h}_q)\Big) \bigg), 
\end{equation}
where $\beta_{pq}$ is the attention coefficient of item $p$ to another item $q$ in the session $S$. $\mathbf{x} \in \mathbb{R}^d$ and $\mathbf{W}_2, \mathbf{W}_3 \in \mathbb{R}^{d \times d}$ are weight matrices. $\mathbf{h}_s$ is the session embedding that contains the session graph structure, temporal information, and user's current intent, so we can use $\mathbf{h}_s$ to infer the user's next interaction in our next step.

\subsection{Hyperbolic evolutionary loss}
Here we introduce how to leverage evolutionary loss to provide recommendations given a specific timestamp. Unlike other works \cite{kumar2019predicting, li2020dynamic}, our evolutionary loss is also fully hyperbolic.
\subsubsection{Evolution formulas}
\label{evol_form}
The core idea of evolutionary loss is to predict the future session and next-item embeddings given a future timestamp and then make recommendations. The prediction results of evolutionary loss do not rely on the sequences like RNN-based models\cite{hidasi2015session, li2017neural} but are based on the final embeddings learned by the TA-HGAT.

As $\mathbf{h}_s$ is the predicted session embedding in the future, we also need an estimated future session embedding to measure whether the predicted embedding is accurate. Assume that the growth of the session embedding is smooth. The embedding vector of the session evolves in a contiguous space. Therefore, we devise a projection function to infer the future session embedding based on the element-wise product of the previous embedding and the time interval. The embedding projection of session $S$ after current time $t$ to the future time $t^+$ is defined as follows:
\begin{align}
&\mathbf{\widehat{h}}_s^{t^+} =\sigma\Big(\mathbf{h}_s^{t}\odot(\mathbf{1} \oplus \mathbf{h}_{t'}) \Big),
\end{align}
where $\mathbf{1} \in \mathbb{R}^d$ is a vector with all elements $1$ and $\odot$ is M\"obius element-wise product. $\mathbf{h}_{t'}$ is the time interval vector, which is learned in the same way as Eq. \ref{interval}. The $\mathbf{1}$ vector is to provide the minimum difference between the last and next session embeddings. With this projection function, the future session embedding grows in a smooth trajectory w.r.t. the time interval.

After learning the projected embedding $\mathbf{\widehat{h}}_s^{t^+}$ of the session $S$, the next step is to apply another projection function to generate the future embedding of the next item $v$, which is denoted as $\mathbf{\widehat{h}}_v^{t^+}$. The projected future item embedding is composed of three components: the projected session embedding, the last item embedding, and the time interval, which are learned in the previous steps. Here, we define the projection formula of next item $v$ as
\begin{align}
&\mathbf{\widehat{h}}_v^{t^+} =\sigma_v\Big( (\mathbf{W}_4 \otimes \mathbf{\widehat{h}}_s^{t^+} ) \oplus (\mathbf{W}_5 \otimes \mathbf{h}_{v_n}) \oplus \mathbf{h}_{t'}\Big),
\end{align}
where $\mathbf{W}_4$ and $\mathbf{W}_5$ denote the weight matrix.

\subsubsection{Loss function}

With the above projection functions, we can achieve the estimated future embeddings of the session and the next item. They are utilized as ground truth embeddings in our loss function. To train the model, the loss function is designed to minimize the distances between model-generated embeddings $\mathbf{h}_s^{t}$, $\mathbf{h}_{v_n}$ and estimated ground truth embeddings $\mathbf{\widehat{h}}_s^{t^+}$, $\mathbf{\widehat{h}}_v^{t^+}$ at each interaction time $t$. Also, another constraint for the item embeddings is necessary to avoid overfitting. We constrain the distance between the embeddings of the most recent two items $v_{n-1}$ and $v_n$ to ensure the last item embeddings are consistent with the previous one. This constraint assumes that the last and next items reflect similar user intent, and the session embedding tends to be stable in a short time. Finally, the loss function is as follows:
\begin{equation}
\begin{split}
\mathcal{L} = \sum_{(s,v,t)\in \{S_i\}_{i=0}^I}& d(\mathbf{\widehat{h}}_v^{t^+},\mathbf{h}_{v_n} ) \oplus \Big( \lambda_s \otimes d(\mathbf{\widehat{h}}_s^{t^+}, \mathbf{h}_s^{t}) \Big) \oplus \\
& \Big( \lambda_v \otimes d(\mathbf{h}_{v_n},\mathbf{h}_{v_{n-1}}) \Big),
\end{split}
\end{equation}
where $\{S_t\}_{i=0}^I$ denotes all sessions in the datasets, and $\lambda_s$ and $\lambda_v$ are smooth coefficients, which are used to prevent the embeddings of the session and items from deviating too much during the update process. $d(\cdot)$ is the hyperbolic distance function which is described in Eq. \ref{distance}.


To make recommendations for a user, we calculate the hyperbolic distances between the predicted item embedding obtained from the loss function and all other item embeddings. Then the nearest top-$k$ items are what we predict for the user.

Compared with traditional BPR loss \cite{rendle2012bpr}, the evolutionary loss is more suitable for time-aware recommendations because it takes time intervals into account. As a result, the changing trajectories are modeled by this loss \cite{kumar2019predicting}, and it can make more precise recommendations for the next item given a specific timestamp.

\section{Experiments}
In this section, we describe the experimental results on two public datasets and compare our proposed TA-HGAT with ten state-of-the-art baseline models. Our experiments are designed to solve the following research questions:
\begin{itemize}
    \item \textbf{RQ1}: How does TA-HGAT compare with other state-of-the-art session-based recommendation models?
    \item \textbf{RQ2}: How do the two modules of time-aware hyperbolic attention, i.e., hyperbolic self-attention with time intervals and hyperbolic soft-attention with users’ current interests, affect the performance of TA-HGAT?
    \item \textbf{RQ3}: How does the hyperbolic evolutionary loss compare with other loss functions?
    \item \textbf{RQ4}: How is the influence of different hyper-parameters, i.e. embedding dimensions?
\end{itemize}

\subsection{Experiment settings}
\subsubsection{Datasets}
We conduct our experiments on two widely used public datasets: Yoochoose and Diginetica. The statistics of these datasets are listed in Table \ref{tab:dataset}.
\begin{itemize}
    \item \textbf{Yoochoose\footnote{https://www.kaggle.com/datasets/chadgostopp/recsys-challenge-2015}} is a public dataset released by the RecSys Challenge 2015, which contains click streams from yoochoose.com within 6 months.
    \item \textbf{Diginetica\footnote{https://competitions.codalab.org/competitions/11161}} is obtained from the CIKM Cup 2016. We use the item categories to initialize the item embeddings.
\end{itemize}

\begin{table}[]
    \centering
    \caption{The number of items, training sessions, testing sessions, the average length, and clicks for each dataset.}
    \resizebox{0.47\textwidth}{!}{\begin{tabular}{cccccc}
    \toprule
         Datasets & Items & train sessions & test sessions & Avg. len & clicks  \\
    \midrule
        Diginetica & 43,097 & 719,470 & 60,858 & 5.12 &  982,961 \\
        Yoochoose1/64 & 16,766 & 369,859 &  55,898 & 6.16 & 557,248 \\
        Yoochoose1/4 &  29,618 & 5,917,746 & 55,898 &  5.71 & 8,326,407\\
    \bottomrule
    
    \end{tabular}}
    
    \label{tab:dataset}
\end{table}

\subsubsection{Evaluation Metrics}
We evaluate the performance of our model with Mean Reciprocal Rank (MRR@K) and Precision (P@K) in the comparison experiments.

\textbf{MRR@K} considers the position of the target item in the list of recommended items. It is set to 0 if the target item  is not in the top-$k$ of the ranking list, or otherwise is calculated as follows:
\begin{equation}
    MRR@K = \frac{1}{N} \sum_{i=1}^N \frac{1}{Rank(v_t)},
\end{equation}
where $v_t$ is the target item and $N$ is the number of test sequences in the dataset.

\textbf{P@K} measures whether the target item is included in the top-$k$ list of recommended items, which is calculated as
\begin{equation}
    P@K = \frac{n_hit}{N}
\end{equation}

\subsubsection{Implementation}
Our model \footnote{The datasets and codes will be available after accepted} is implemented with PyTorch 1.12.1 \cite{paszke2019pytorch} and CUDA 10.2. In the testing phase, we take the interval between the session's last timestamp and the testing item's timestamp as a part of the input to obtain the recommendation list. This setting is different from other baseline models as they cannot deal with temporal information. In fact, this setting meets the actual situation in the industry because our model can provide recommendations as soon as the user logs into the website, and we can easily obtain the real-time time interval.

\subsection{Performance comparison (RQ1)}
To demonstrate the effectiveness of TA-HGAT, we conduct experiments on two public datasets and compare the model with ten state-of-the-art baseline models. 
\subsubsection{Baseline models}
\begin{itemize}
    \item \textbf{S-POP} takes the most popular items of each session as the recommended list.

    \item \textbf{FPMC}\cite{rendle2010factorizing} is a Markov chain-method for sequential recommendation, which only takes the item sequences in session-based recommendation since user features are unavailable.
    
    \item \textbf{GRU4REC}\cite{hidasi2018recurrent} is the first work that applies RNN to the session-based recommendation to learn the sequential dependency of items.

    \item \textbf{NARM}\cite{li2017neural} utilizes an attention mechanism to model the sequential behaviors and the user’s primary purpose with global and local encoders.

    \item \textbf{STAMP}\cite{liu2018stamp} employs an attention and memory mechanism to learn the user's preference and takes the last item as recent intent in the session to make recommendations.
    
    \item \textbf{SR-GNN}\cite{wu2019session} is the first work that model a session into a graph. It resorts to the gated graph neural networks to learn the complex item transitions in the sessions.
    
    \item \textbf{TAGNN} \cite{yu2020tagnn} improves SR-GNN by learning the interest representation vector with different target items to improve the performance of the model.
    
    \item \textbf{NISER+} \cite{gupta2019niser} handles the long-tail problem in SBR with L2 normalization and dropout to alleviate the overfitting problem.
    
    \item \textbf{SGNN-HN} \cite{pan2020star} applies a star graph neural network to consider the items without direct connections.
    
    \item \textbf{HCGR} \cite{guo2021hcgr} models the session graphs in hyperbolic space and makes use of multi-behavior information to improve performance. In our experiments, we don't use the behavior information as the datasets didn't provide it and we are modeling a more general scenario.
\end{itemize}

\begin{table}[t]
  \centering
  \caption{Experiments on Diginetica and Yoochoose datasets compare TA-HGAT with ten baseline models based on the top-20 of the ranking list in Mean Reciprocal Rank (MRR@20) and Precision (P@20). The bold and underlined numbers on each dataset and metric represent the best and second-best results, respectively. "Improv." refers to the minimum improvement among all baselines.}
    \resizebox{0.48\textwidth}{!}{\begin{tabular}{clccccccccc}
    \toprule
    {Models}  
  &\multicolumn{2}{c}{Diginetica}  &\multicolumn{2}{c}{Yoochoose 1/64}  &\multicolumn{2}{c}{Yoochoose 1/4}  \\
    \cmidrule(lr){2-3}   \cmidrule(lr){4-5}  \cmidrule(lr){6-7}     
    & MRR@20 & P@20 & MRR@20 & P@20 & MRR@20 & P@20  \\
    \midrule 
      S-POP   &  13.68 & 21.06  &  18.35 &  30.44 & 17.75 & 27.08 \\
      FPMC & 8.92 & 31.55  & 15.01 &  45.62 & - & -  \\
      GRU4REC   &  8.33  &  29.45 &  22.89 &  60.64   & 22.60 &  59.53 \\
      NARM  &   16.17   & 49.70 & 28.63  & 68.32   & 29.23  & 69.73  \\
      STAMP & 14.32 &  45.64 &  29.67 &  68.74 &  30.00 &  70.44  \\
      SR-GNN & 17.59 & 50.73 & 30.94 & 70.57 &  31.89 &  71.36 \\
      TAGNN &  18.03 & 51.31 & 31.12 & 71.02  & 32.03 & 71.51   \\
      HCGR & 18.51 & 52.47 & 31.46 & 71.13  & 32.39 & 71.66   \\
      NISER+ & 18.72 &  53.39 & 31.61 & 71.27 & 31.80 & 71.80 \\
      SGNN-HN  &   \underline{19.45} & \underline{55.67}  & \underline{32.61} & \underline{ 72.06}   & \underline{32.55} & \underline{72.85} \\
      
      TA-HGAT &   \textbf{19.73} & \textbf{56.28}  & \textbf{32.90} & \textbf{72.75} & \textbf{32.94} & \textbf{73.56} \\
      \midrule
      Improv. & 1.44\% & 1.10 \% &  0.89\% & 0.96\% & 1.20\% & 0.97\% \\ 
    \bottomrule
    \end{tabular}}
  \label{table:performance}
\end{table}

\subsubsection{Result analysis}
The complete experimental results of the comparison study are shown in Table \ref{table:performance}. From the results, we have the following observations:
\begin{itemize}
    \item Our proposed TA-HGAT outperforms all baseline models on all datasets and metrics, which demonstrates the effectiveness of the model. Besides, HCGR, which is another hyperbolic graph-based SBR model, has achieved better performance than the graph-based SBR model SR-GNN but worse than our model. Compared to HCGR, our model improves 4.3\% and 4.1\% on average over three datasets on metrics MRR@20 and P@20, respectively. HCGR is better than SR-GNN, indicating that hyperbolic embeddings match session graphs. And the improvement of TA-HGAT over HCGR shows the importance of temporal information in the SBR task.
    \item In Table \ref{table:performance}, we also observe that our model has a better performance on dataset Diginetica than Yoochoose. On average, the performance of TA-HGAT on Diginetica outperforms Yoochoose for 37.8\% and 14.0\% on metrics MRR@20 and P@20, respectively. This phenomenon may result from the initial features of items.  In Diginetica, each item has its category label, and we transform this feature into a one-hot vector as the initial embedding of the item. In HCGR, we model the initial feature to a feature vector in the hyperbolic space, which is shown in Eq. \ref{initial} and \ref{vector_multiply}. Differences in performance between Diginetica and Yoochoose indicate that the hyperbolic embeddings have a better expression ability on the item features.
\end{itemize}

\subsection{Ablation study (RQ2)}
In the TA-HGAT, we have two main modules in time-aware hyperbolic attention: hyperbolic self-attention with time intervals and hyperbolic soft-attention with users’ current interests. In this section, we evaluate their effectiveness separately to show the improvement compared with the ablation models without these two modules.

\begin{figure*}[tbp]
\centering
\subfigure[Diginetica]{
\includegraphics[width=1.75in]{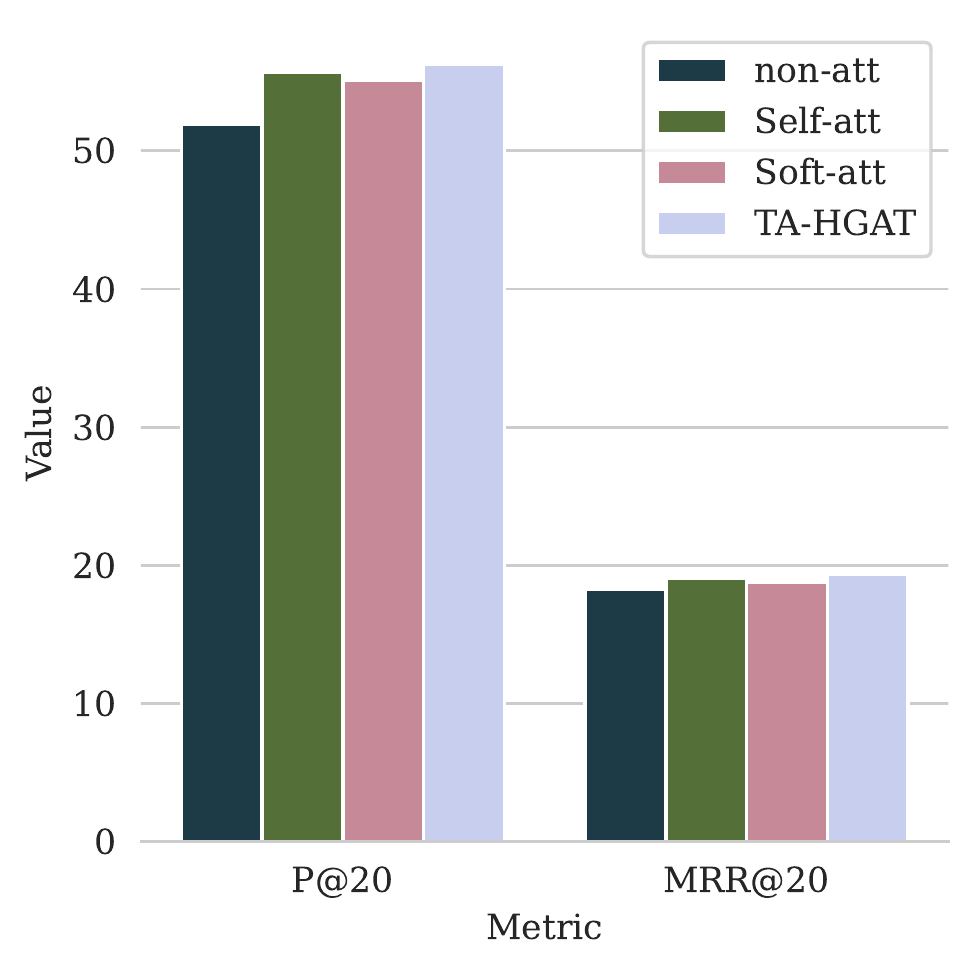}	
}
\quad
\quad
\subfigure[Yoochoose 1/64]{
\includegraphics[width=1.75in]{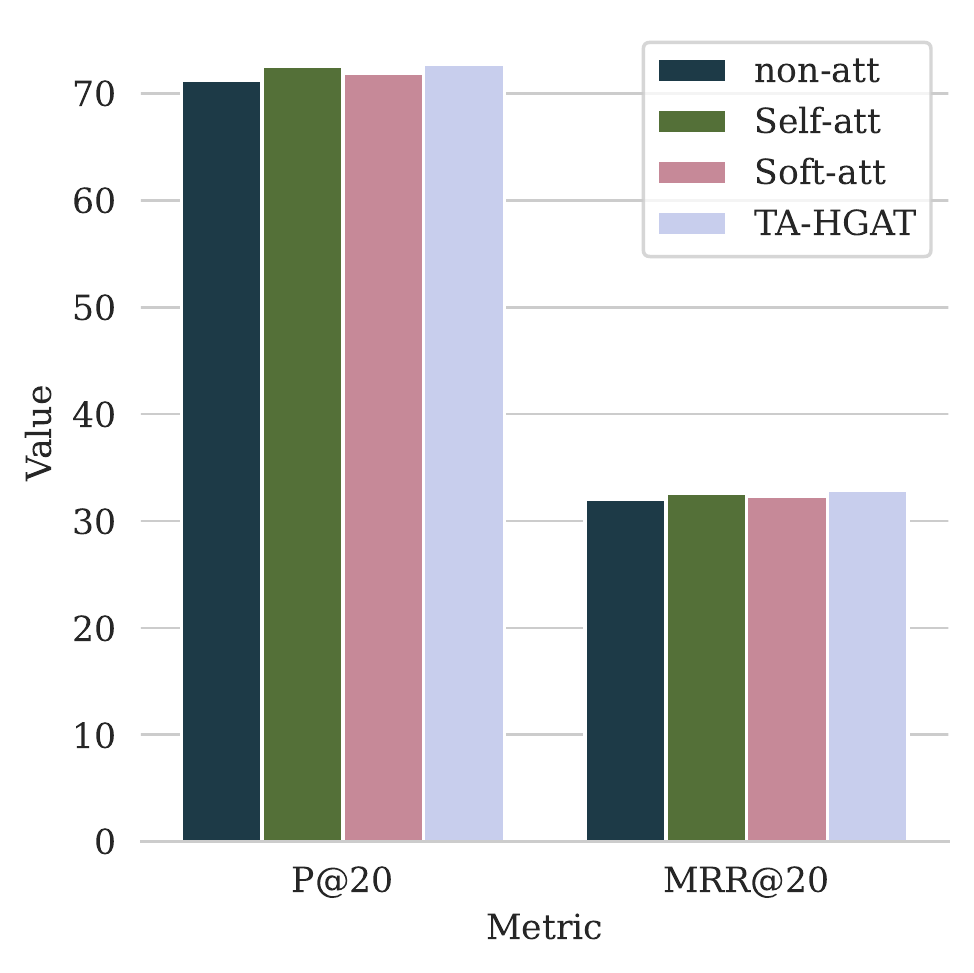}	
}
\quad
\quad
\subfigure[Yoochoose 1/4]{
\includegraphics[width=1.75in]{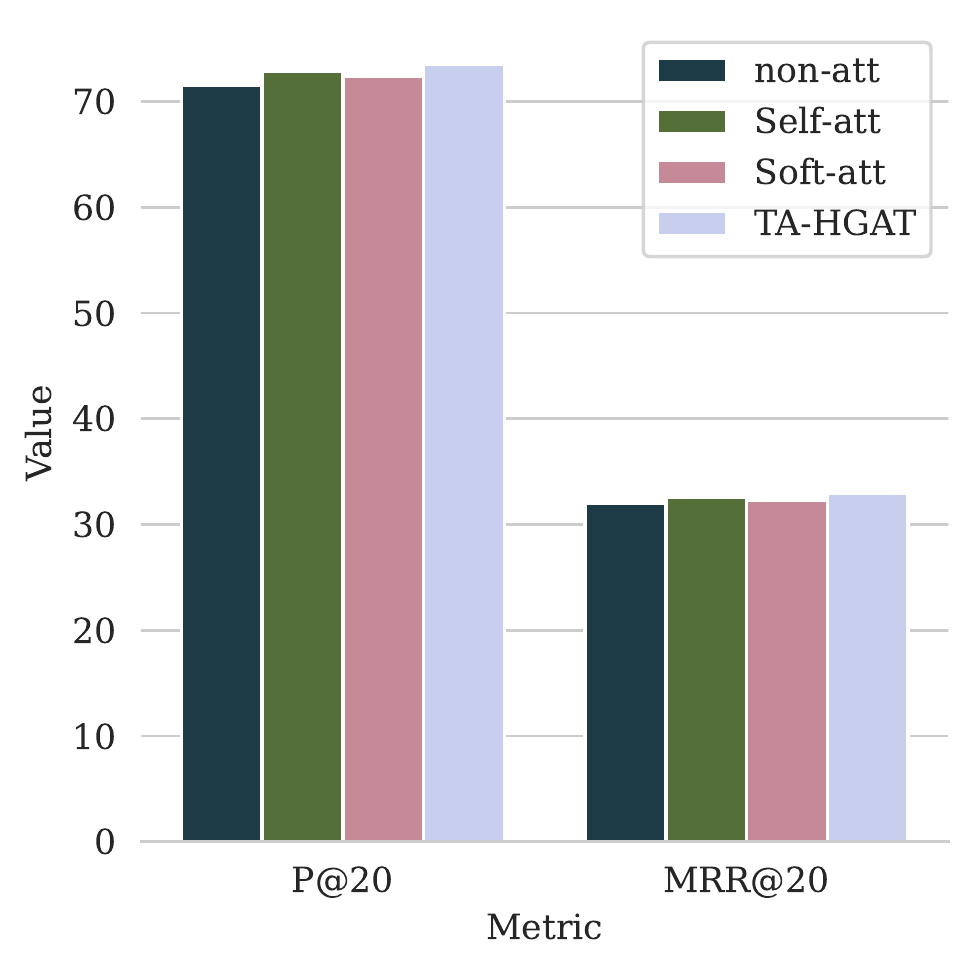}	
}
\caption{The ablation study of TA-HGAT. 'non-att' is our model without attention layers. 'Self-att' and 'Soft-att' are composed of only self-attention and soft-attention layers, respectively. TA-HGAT is the complete model. }
\label{fig:ablation}
\end{figure*}

We set up four separate ablation models to compare the effectiveness of each attention layer. The first ablation model is \textit{no-att}, in which we remove both the attention layers and only conduct the aggregation operations directly. The second and third ones are \textit{self-att} and \textit{soft-att}, and these two ablation models only include the self-attention and soft-attention layers, respectively. The fourth one is \textit{TA-HGAT}, which is the complete model. The comparison results of the ablation models on datasets Diginetica and Yoochoose are illustrated in Figure \ref{fig:ablation}.

From Figure \ref{fig:ablation}, we observe the following results:
\begin{itemize}
    \item On both Diginetica and Yoochoose datasets, the non-att performs worst, and TA-HGAT performs best. The results show the effectiveness of the attention layers. This is because the TA-HGAT makes full use of the temporal information. Compared to the GNNs without temporal information, our model builds the relations between items with time intervals and also considers users' current interests. Hence, the rich information helps the model to achieve better results.
    
    \item Self-att performs better than soft-att, which means time intervals are relatively more meaningful than users' current interests. This phenomenon may be due to the fact that users' current interests are more complicated, so the last item cannot fully represent them. In contrast, the time interval is a more straightforward feature, so our proposed hyperbolic self-attention layer can handle this information effectively.
\end{itemize}

\subsection{Comparison of loss functions (RQ3)}
In this section, we compare our proposed hyperbolic evolutionary loss to conventional loss functions, i.e., BPR \cite{rendle2012bpr} and softmax loss \cite{wu2019session}. Because the learned session embeddings in the output of our model are in the hyperbolic space, we need to use the logarithmic map to project the embeddings back to the Euclidean space before applying BPR and softmax loss.

The comparison results are shown in Table \ref{table:loss}. The hyperbolic evolutionary loss is denoted as TA-HGAT in the table. From this table, we can find that the performance of BPR and softmax loss is similar, but our proposed hyperbolic evolutionary loss has a clear improvement compared to the other losses. This observation demonstrates that considering the specific timestamp is effective for the SBR task models designed in hyperbolic space. 

\begin{table}[t]
  \centering
  \caption{Comparison of performance for different loss functions.}
    \resizebox{0.48\textwidth}{!}{\begin{tabular}{clccccccccc}
    \toprule
    {Loss}  
  &\multicolumn{2}{c}{Diginetica}  &\multicolumn{2}{c}{Yoochoose 1/64}  &\multicolumn{2}{c}{Yoochoose 1/4}  \\
    \cmidrule(lr){2-3}   \cmidrule(lr){4-5}  \cmidrule(lr){6-7}     
    & MRR@20 & P@20 & MRR@20 & P@20 & MRR@20 & P@20  \\
    \midrule 
      Softmax   &  19.38 & 55.81  &  32.67 &  72.10 & 32.72 & 72.95 \\
      BPR & 19.43 & 55.97  & 32.53 &  72.28 & 32.66 & 72.84  \\
      TA-HGAT &   \textbf{19.73} & \textbf{56.28}  & \textbf{32.90} & \textbf{72.75} & \textbf{32.94} & \textbf{73.56} \\
    \bottomrule
    \end{tabular}}
  \label{table:loss}
\end{table}

\subsection{Hyperparameter analysis (RQ4)}
\begin{figure}[tbp]
\centering
\subfigure[Diginetica MRR]{
\includegraphics[width=1.55in]{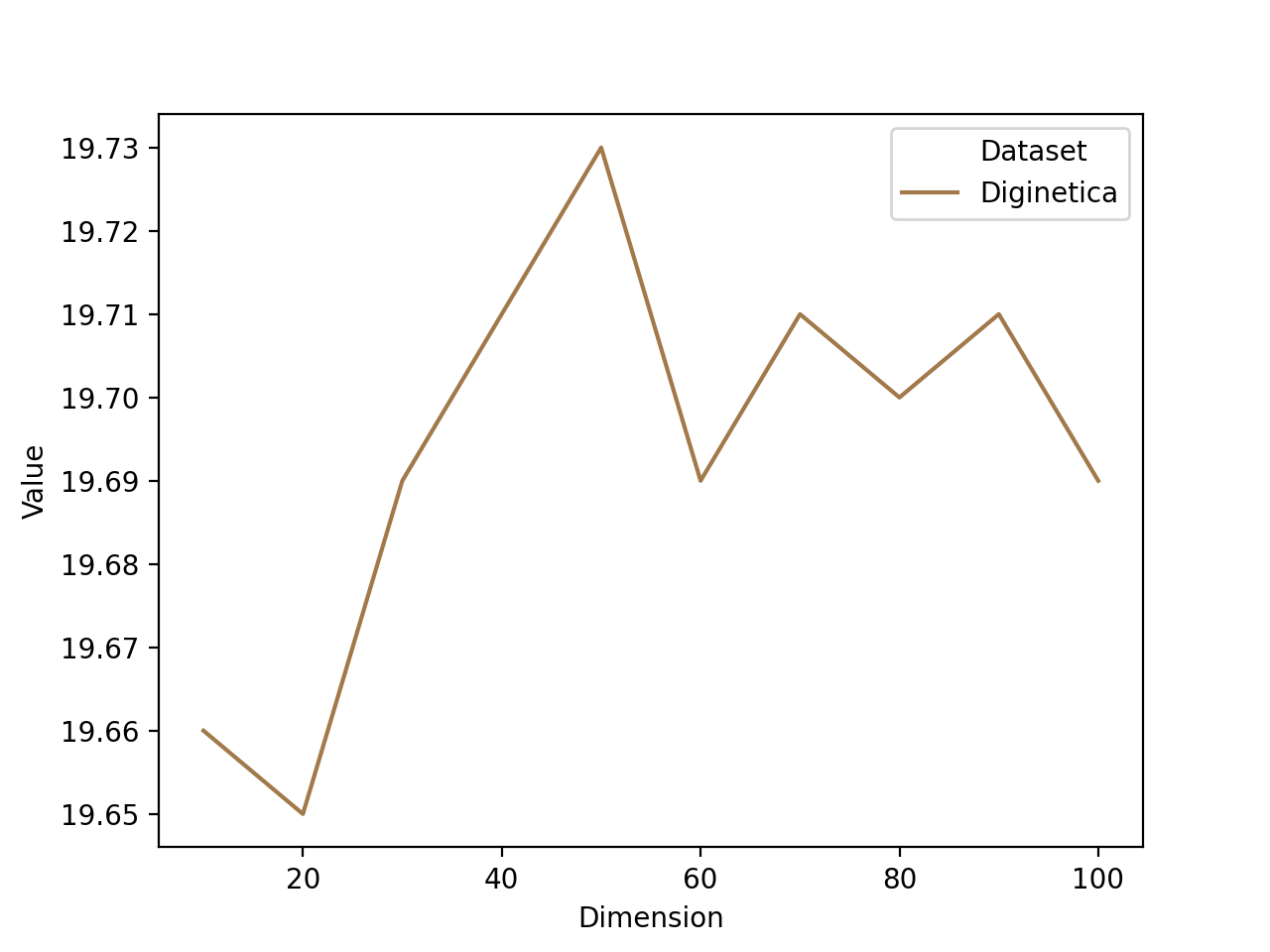}	
}
\quad
\subfigure[Yoochoose MRR]{
\includegraphics[width=1.55in]{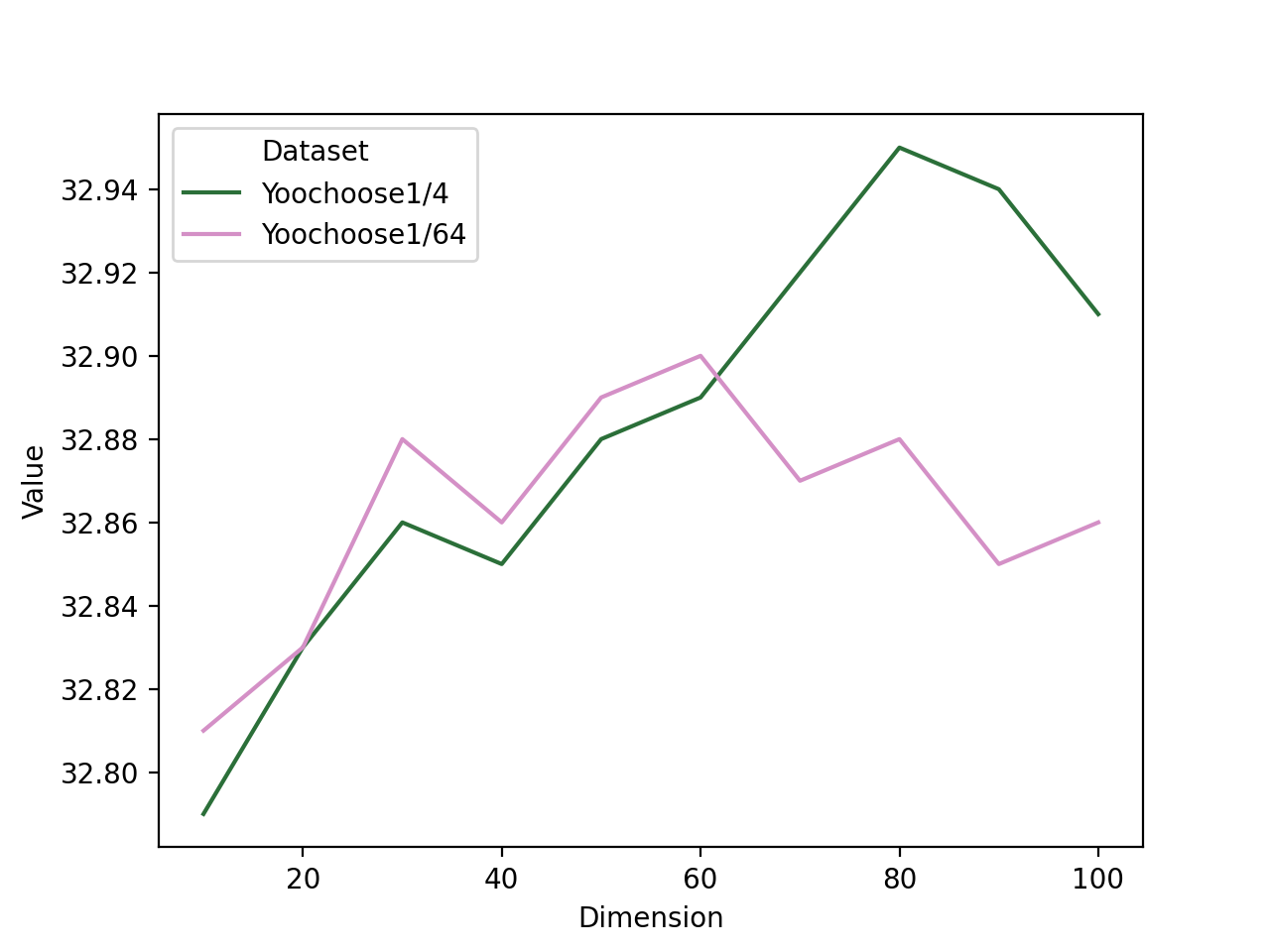}	
}
\quad
\subfigure[Diginetica Precision]{
\includegraphics[width=1.55in]{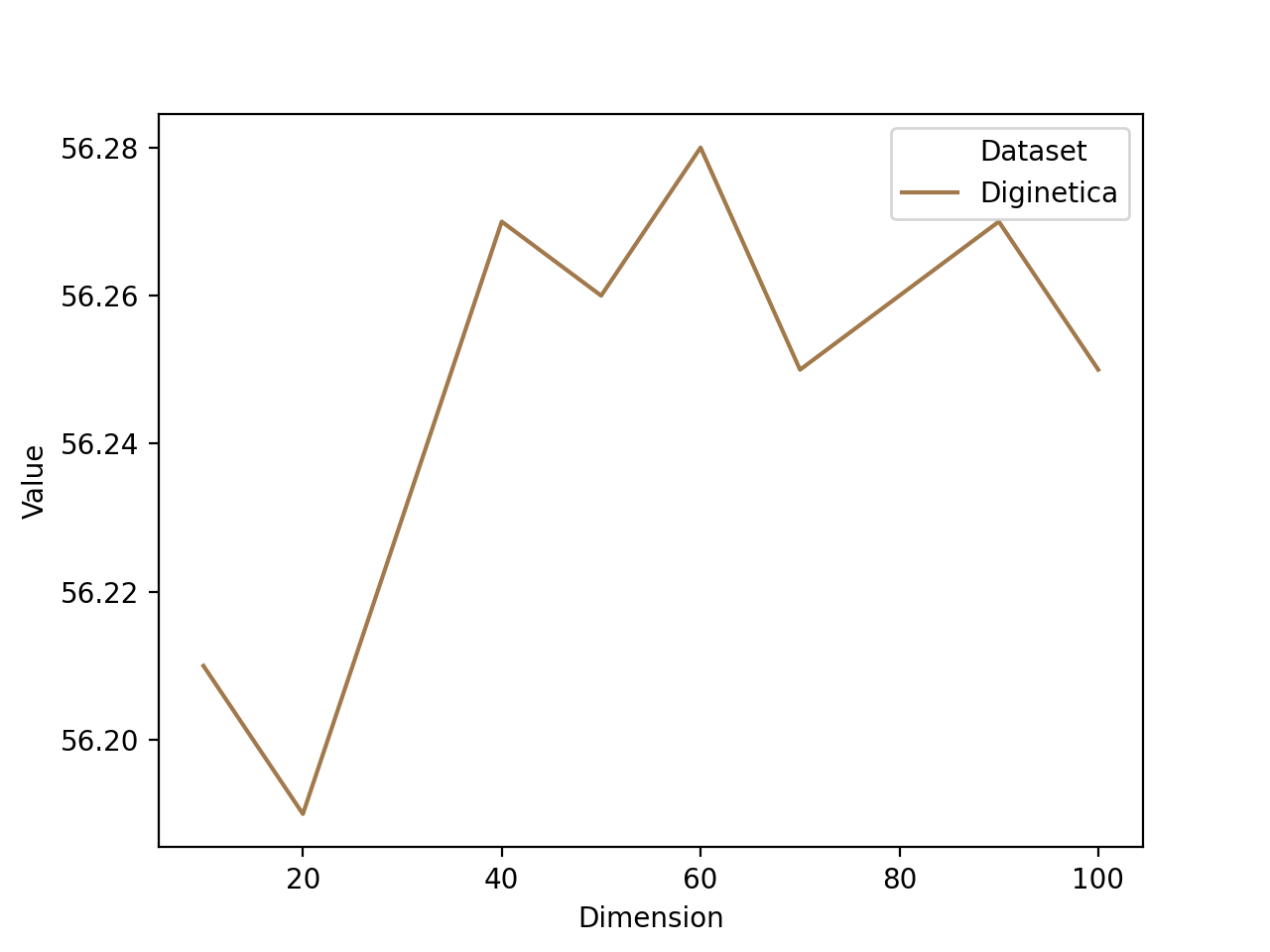}	
}
\quad
\subfigure[Yoochoose Precision]{
\includegraphics[width=1.55in]{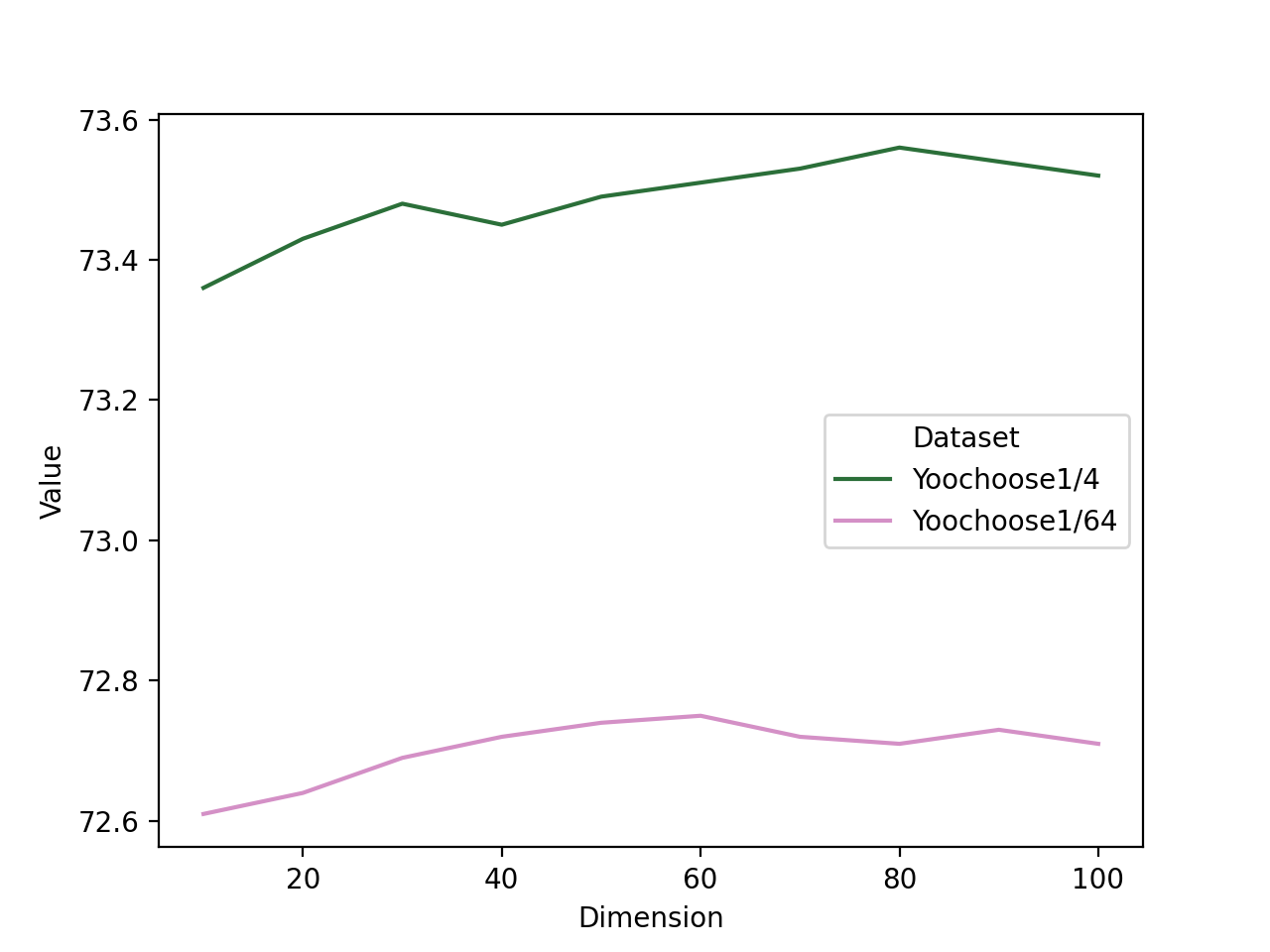}	
}
\caption{The hyperparameter analysis of the embedding dimensions.}
\label{fig:hyperparameter}
\end{figure}

The embedding dimension is the hyperparameter in our proposed model, so we test the influence of different embedding dimensions in this section. The embedding dimensions range from 20 to 100. The results of the hyperparameter analysis are illustrated in Figure \ref{fig:hyperparameter}.

It is observed that a proper embedding dimension is essential for learning the item and session representations. From Figure \ref{fig:hyperparameter}, we can see that the Diginetica and Yoochoose 1/64 all achieve the best performance when the embedding dimension is 60, and the best result of Yoochoose 1/4 is 80. Because Yoochoose 1/4 is much larger than the other two datasets, it indicates that larger datasets need larger embedding space.

\section{Related works}
\subsection{Hyperbolic spaces}
Recent research has shown that many types of complex data exhibit a highly non-Euclidean structure\cite{bronstein2017geometric}. In many domains, e.g., natural language \cite{tifrea2018poincar}, computer vision\cite{yau1983hierarchical}, and healthcare\cite{wilkens2005hiers}, data usually has a tree-like structure or can be represented hierarchically. Since this type of data contains an underlying hierarchical structure, capturing such representations in Euclidean space is difficult. To solve this problem, current studies are increasingly attracted by the idea of building neural networks in Riemannian space, such as the hyperbolic space, which is a homogeneous space with constant negative curvature \cite{nickel2017poincare}. Compared with Euclidean space, hyperbolic space in which the volume of a ball grows exponentially with radius instead of growing polynomially. Because of its powerful representation ability, hyperbolic space has been applied in many areas. 
For instance, \cite{dhingra2018embedding} learns word and sentence embeddings in hyperbolic space in an unsupervised manner from text corpora. \cite{khrulkov2020hyperbolic} demonstrates that hyperbolic embeddings are beneficial for visual data. \cite{chami2019hyperbolic} proposes Hyperbolic Graph Convolutional Neural Networks, which combines the expressiveness of GCNs and hyperbolic geometry to learn graph representations. These works show the potential and advantages of hyperbolic space in learning hierarchical structures of complex data.

Based on the performance of hyperbolic space in these fields, it is natural for researchers to think of applying hyperbolic learning to recommender systems. \cite{chamberlain2019scalable} justifies the use of hyperbolic representations for neural recommender systems. \cite{vinh2020hyperml} proposes HyperML to bridge the gap between Euclidean and hyperbolic geometry in recommender systems through a metric learning approach. \cite{sun2021hgcf} proposes a hyperbolic GCN model for collaborative filtering. \cite{wang2021hypersorec} presents HyperSoRec, a novel graph neural network (GNN) framework with multiple-aspect learning for social recommendation. 

\subsection{Session-based Recommendation}
Session-based Recommendation (SBR) has increasingly engaged attention in both industry and academia due to its effectiveness in modeling users' current interests. In the recent SBR studies, there are mainly three types of methods that apply deep learning to SBR and have achieved state-of-the-art performance, which are sequence-based \cite{hidasi2015session, tan2016improved}, attention-based \cite{li2017neural, liu2018stamp} and graph-based models \cite{wu2019session}. GRU4REC \cite{hidasi2015session} is the most representative work in the sequence-based SBR models. It employs GRU, a variant of RNN, to model the item sequences and make the next-item prediction. Following GRU4REC, some other papers \cite{tan2016improved, quadrana2017personalizing, hidasi2018recurrent} improve it with data augmentation, hierarchical structure, and top-$k$ gains. Attention-based models aim to learn the importance of different items in the session and make the model focus on the important ones. NARM \cite{li2017neural} utilizes an attention mechanism to model both local and global features of the session to learn users' interests. STAMP \cite{liu2018stamp} combines the attention model and memory network to learn the short-term priority of sessions. Graph-based models connect the items in a graph to learn their complex transitions. SR-GNN \cite{wu2019session} is the first work that models the sessions into graphs. It leverages the Gated Graph Neural Network (GGNN) to model the session graphs and achieve state-of-the-art performance. Based on SR-GNN, \cite{xu2019graph, yu2020tagnn} improve SR-GNN with attention layers. \cite{qiu2019rethinking, chen2020handling} consider the item order in the session graph in the model. \cite{wang2020global, liu2021case4sr, pang2022heterogeneous} take additional information such as global item relationship, item categories, and user representations into account to design more extensive models. However, all these methods fail to consider the hierarchical geometry of the session graphs and the temporal information.

\subsection{GNN-based Recommendation Models}
GNNs have proven to be useful in different research fields \cite{peng2019fine, liu2021medical, dou2020enhancing, liu2020heterogeneous, li2021higher, hei2021hawk}. There also exist many works considering the graph structures in data modeling of recommender systems. Generally, there are two main ways regarding the graph structure and embedding space. One way is to model the user-item interaction graph in Euclidean spaces. Among them, \cite{zheng2018spectral, berg2017graph, yu2020graph} perform graph convolution on the user-item graph to explore their interactions. \cite{wang2019neural, he2020lightgcn, liu2020basket} utilize layer-to-layer neighborhood aggregation in GNNs to capture the high-order connections. \cite{li2021pre} pre-trains user-user and item-item graphs separately to learn the initial embeddings of the user-item interaction graph. \cite{li2020dynamic} models the changes in user-item interactions with a dynamic graph and evolutionary loss. These works apply GNN to learn from high-dimensional graph data and generate low-dimensional node embeddings without feature engineering, but the learned embeddings are all in Euclidean spaces, while some graph data may be more suitable to other geometries in the representation learning.

The other way is to model the recommendation graph in hyperbolic space to learn the hierarchical geometry. HGCF \cite{sun2021hgcf} applies hyperbolic GCN to learn the node embeddings using a user-item graph. Wang et al. \cite{wang2021fully} propose a fully hyperbolic GCN where all operations are conducted in hyperbolic space. Xu et al. \cite{xu2020product} model the product graph in a knowledge graph and learn the node embeddings in hyperbolic space. HCGR \cite{guo2021hcgr} is a novel hyperbolic contrastive graph representation learning method to make session-based recommendations. None of these models utilize the time-relevant information in the session graphs to improve the recommendation accuracy.
In this paper, we propose a novel framework incorporating a time-aware graph attention mechanism in hyperbolic space, which is specifically devised for the session-based recommendation.


\section{Conclusion}
Session-based Recommendation (SBR) is to predict users’ next interested items based on their previous sessions. Existing works model the graph structure in the sessions and have achieved state-of-the-art performance. However, they fail to consider the hierarchical geometry and temporal information in the sessions. In this paper, we propose TA-HGAT, a hyperbolic GNN-based model that considers the time interval between items and users' current interests. Experiment results demonstrate that TA-HGAT outperforms other SBR models on two real-world datasets.

For future work, we will extend our model to more general recommender systems. The time intervals are not only in the SBR problem, but also in almost all recommender systems. As a result, we want to test how our model performs on other recommendation problems, e.g., next-basket recommendation and point-of-interest recommendation, where temporal information plays a crucial role in providing recommendations.

\section{Acknowledgement}
This work is supported in part by NSF under grants III-1763325, III-1909323, III-2106758, and SaTC-1930941.

\bibliographystyle{IEEEtran}
\balance
\bibliography{ref} 

\end{document}